\documentclass[reprint,
 amsmath,amssymb,
 aps,
pra,
]{revtex4-2}
\usepackage{quoting}
\usepackage{physics}
\usepackage{graphicx}
\usepackage{dcolumn}
\usepackage{bm}
\usepackage[caption=false]{subfig}
\usepackage{lineno}
\usepackage{xcolor}

\begin{document}


\title{Chirp asymmetry in Zeeman electromagnetically induced transparency}
+
\author{Joseph Gorkos}
 \email{jwgorkos@student.ysu.edu}
\author{Karsten Grenzig}
 \email{kagrenzig@student.ysu.edu}
 \author{Erfan Nasirzadeh Orang}
 \email{enorang@student.ysu.edu}
 \author{Victoria Thomas}
 \email{victorialynthomas@gmail.com}
 \author{Declan Tighe}
 \email{detighe@student.ysu.edu}
\author{Michael Crescimanno}
\affiliation{Department of Physics, \\Youngstown State University, Youngstown Ohio}
\email{dcphtn@gmail.com}

\date{\today}

\begin{abstract}
The simplest three-level system exhibiting electromagnetically induced transparency (EIT) exhibits an effective conjugation symmetry as well as a permutation symmetry. Breaking conjugation symmetry leads to a distinct chirp asymmetry, {\it i.e.} the differential response to a frequency increase versus a frequency decrease. Hanle-Zeeman EIT resonance is an ideal platform for testing the theory of chirp asymmetry because so many optical parameters of the system can be changed experimentally. We describe the theory and compare it to an experiment using $^{87}$Rb in a buffer gas cell. In contrast with earlier multi-photon chirp asymmetry work this present effort explores the asymmetry at nearly one billionth the earlier chirp rate, yet displays its universal features. Chirp asymmetry may have metrological consequences for understanding systematic dependence on modulation/demodulation parameters. 
\end{abstract}

\maketitle
 
\section{Introduction}
\label{sec:intro} 
Due to a myriad of physical effect, optical resonances typically depart from the simple Lorentz-Schwinger-Weisskopf form. Beyond naive center shifts, lineshape distortions can include effects that (at least approximately) respect the underlying symmetric nature of that form (for example, Doppler effects, far off-resonant collisional broadening, power broadening in lowest order and magnetic effects at weak fields) and those that generate lineshape asymmetries (including AC stark effects in other non-resonant light fields, detunings in multi-photon processes, some inhomogeneous broadening effects). Minimizing lineshape asymmetries is important for precision metrology of various types \cite{nadezhinskii, brown, truong}, leading, for one example, to modulation scheme-dependence in line center measurements \cite{phillips, yudin}. 

Lineshape asymmetries in single photon transitions also lead to chirp asymmetries, {\it i.e.} a difference in the optical response to a frequency up-chirp versus a down-chirp. A recent systematic study of chirp asymmetry (see \cite{commons}) using saturation spectroscopy emphasized its connection to the general theory of transient response in systems with broken discrete symmetries. Among the universal behavior of such systems is a chirp asymmetry that varies initially linearly with chirp speed and symmetry breaking parameter and, further, that saturates (to a value not necessarily $\pm 1$) at high chirp speeds and parameter. This behavior is an example of the Sakharov conditions \cite{sakharov91} that, for example, figure prominently in models of early universe leptogenesis. 

In earlier work (\cite{commons}) the lineshape asymmetry was fixed by the level structure of the excited state hyperfine manifold of the Rubidium atom. There the asymmetry of the optical response was measured as a function of the chirp speed, yielding the characteristic linear growth at small chirp speeds followed by an asymmetry plateau at high chirp speeds. However the precise connection between the slope of the linear part of this asymmetry curve and the C-symmetry breaking parameter was not testable in that experiment. Furthermore in that approach a study of the chirp asymmetry's dependence on linewidth and non-linear optical broadening were likewise unavailable. 

In light of the forgoing a more detailed test of this potentially universal theoretical/phenomenological description of chirped response would entail being able to observe, for example, systematic changes in the later as one changed the value of the symmetry breaking parameter. Additionally it would be instructive to experimentally test chirp asymmetry's dependence on the resonance width since that parameter plays a role in quantifying how far out of equilibrium has been driven the system at fixed chirp rate. It would also be useful to use an experimental protocol for insight into the roles played by each of the intrinsic width and the contributions to the observed width from homogeneous and inhomogeneous effects. In chirped multiphoton processes higher-order optical effects such as AC Stark effects are implicated in playing an important role, but the precise connection of such contributions to the underlying Sakharov framework for this asymmetry is unclear. Note also that chirp asymmetry phenomenology for quantum system is expected to be richer than that of early universe leptogenesis models in part due to the fact that the former receives modulations from both coherences and populations, whereas the later does not. 

Our goal here is to demonstrate and amplify the chirp asymmetry systematics in part by addressing the limitations of prior studies in sweep speed, detection bandwidth and lack of tunability of the symmetry breaking parameters. We do this by locating the phenomenon in a 
Zeeman electromagnetically induced transparency resonance. One utility of using this resonance follows in part from the relative ease which one can experimentally control the overall lineshape asymmetry (by changing a detuning) and the resonance width, both from homogeneous and inhomogeneous sources. A longitudinal magnetic field gradient is used to increase the inhomogeneous broadening of the resonance whereas power broadening controllably contributes to the homogeneous broadening of the line. We also note that the (two-photon-) chirp rates we employ in this experiment are nearly one billion times smaller than in the previous study, and we  interrogate a completely different set of atomic transitions and processes. This highlights the universality of the chirp asymmetry phenomenology and provides a more rigorous test of the theory of chirp asymmetry and its connection to broken discrete symmetries in multi-photon quantum optics.

Studying chirp asymmetry in EIT may lead to a simple protocol for determining and potentially ameliorating subtle lineshape asymmetries that directly affect precision spectroscopic  measurements. We note that the asymmetry we are studying here has been long noted, if, however not critically studied for its connection to a broken discrete symmetry and the Sakharov conditions. For one example Ref. \cite{grewal} notes,

\begin{quotation} "However, the more thorough analysis reveals a small asymmetry between two sides of the signal (one arm of the resonance is larger than the other). This is not observed in conventional NMOR (sic. non-linear magneto-optical rotation) signals and indicates incomplete equilibration of the system in Fig. 2(a). It was verified with independent measurements that the asymmetry disappears for even lower sweep rates (not shown). For larger sweep rates ($> 10 \mu$G/s), the asymmetry increases until the signal starts to oscillate [Figs. 2(b) and 2(c)] while crossing B = 0". 
\end{quotation}

The temporal response of Hanle/Zeeman EIT to transients of various types has received significant experimental and theoretical scrutiny \cite{shwa, valente02}, chiefly focussed on the 'switch-on' (or its time reverse 'switch-off') transient response at fixed detuning \cite{chen_hx, li1995, li1996, momeen, nikolic} whose phenomenology is apparently rather different than under a chirp asymmetry protocol (as is its theoretical explication, see, for example, \cite{greentree}). Many of these references also rarely quantitatively compare the 'switch-on' to the 'switch-off'  transient behavior. Those that do include \cite{, jin}, where however the comparison is not connected to any analytical understanding of transient differences due to broken discrete symmetries. For example, in \cite{jin}, the authors do have a graph of the positive and negative chirp behavior but note ''Here, due to the low sweep rate (10 KHz/s), the OMR (sic. optical magnetic resonance) curve still has no difference for the opposite sweep direction.''  In Ref.\cite{park04} hyperfine EIT transients were studied via chirp, but they did not report chirp asymmetry. 
Significant effort \cite{chen_yf,jin} has also been focused on the transient regime that included a pronounced oscillatory character (referred to as the "non-adiabatic regime"). 

Importance of two-photon lineshape asymmetry and its transfer to modulation induced and other clock shifts has also been of long term interest \cite{phillips, yudin}. Chirp asymmetry in two-photon excitation has a storied history (\cite{broers,paspalakis,zhou}), often the two-photon line asymmetry (in leading order) is due to the differential AC stark effect and explicitly breaks the effective conjugation symmetry (see for example  Eq.~(41) of Ref.\cite{brewer, sharaby}) at non-zero one-photon detuning. In this note however, the two-photon lineshape asymmetry is also related to the one-photon detuning but is intrinsic to the EIT with additional AC Stark effects contributing to higher order. Experiments to arrive at a clearer understanding of the contributions of higher order AC Stark effects to the asymmetry are presently underway.   

The connection between this chirp study and the earlier studies of the Landau-Zener (LZ) transition bears mention. Some of the more recent theoretical studies do not report any chirp asymmetry \cite{vasilev, torosov2, shwa, rangelov2} because they entail systems with unbroken discrete symmetries. For example, the usual derivations for the LZ probability depend only on the chirp speed magnitude and not its direction and strikingly has an essential singularity at slow chirp rates. It is noteworthy that Ref.\cite{kitamura} concerns itself with the LZ in a system with a broken discrete symmetry (parity) and shows a dramatically different picture in that the difference in the LZ rates is now proportional to the chirp rate and scales with the amount of symmetry breaking, as indicated in the exposition of the general theory (below).  

In Ref.  \cite{jin}, they study optical transients in the magnetic resonance of the very same transition we study here ($F=2 \xrightarrow{} F'=1$ in $^{87}$Rb vapor), however, they do not systematically study chirp asymmetry and do not connect the transient behavior they study to symmetry/symmetry breaking in the quantum optical system. In particular, the single reported comparison of up- and down- chirped optical response in that reference notes no difference, from the understanding herewith, presumably due to slow chirp rate or tuning other parameters to the symmetry restored values. 

Importantly, space precludes we leave out the connection between chirp asymmetry and non-reciprocal behavior of excursions in a complex order parameter \cite{pack1, pack2}. Connecting broken discrete symmetries with non-trivial closed paths in parameter space as indicated by the forgoing references may be a very useful theoretical framework \cite{ohga} for further amplifying the systematics associated with chirp asymmetry. Since the emphasis of this present work is primarily as a demonstration/application of the basic theory behind this aspect of chirp phenomenology, we likewise do not explore the possible relevant generalizations of the pulse-area theorem \cite{hahn, eberly, zeldovich} to this process, but are currently pursuing both of these avenues in future work. 

Below after reviewing the theory we describe the experimental protocol and finally discuss their quantitative and qualitative comparison.

\section{Theory} 
Although all our data relates to the $^{87}$Rb F=2 $\rightarrow$ F'=1 transition which receive contributions from multiple magnetic sublevels, it will suffice to restrict our theory considerations to a simplified 4 level model Fig.\ref{fig:setup}a. The states $\ket{1}$, $\ket{2}$ and $\ket{3}$ represent different magnetic sublevels (which one can think of as $\pm1$ and 0, resp. of the F=2) and the $\ket{0}$ and $m=0$ excited state (F'=1). 
When the F=2 states are degenerate (no magnetic field) then
under illumination by linearly polarized light the fixed phase relation between the circular polarization eigenstates pumps atoms into the dark state that is a fixed linear superposition of the $\ket{1}$ and $\ket{2}$ and alternatively of course into $\ket{3}$, leading to an increased transmission of the light we identify as EIT (alternatively the Hanle resonance, or Zeeman EIT). However, the presence of even a modest (mG) magnetic field directed along the direction of propagation of the lightfield causes Zeeman splittings between the F=2 states that reduces the system's dark states to just the $|3>$ alone. The Zeeman energy splitting causes differential temporal phase advance,so the former linear superposition of the $\ket{1}$ and $\ket{2}$ which was dark rotates into a light absorbing state on a Larmor timescale. 

A density matrix (semiclassical) Bloch equation encapsulates this phenomenology rather economically. From $\partial_t \rho = -i[H,\rho] + {\cal L}\rho$ we may follow the temporal evolution of the density matrix of the system in the rotating wave approximation (RWA). In addition to Tr($\rho$)=1 and $\rho^\dagger = \rho$, the equation set reads, 
\begin{equation}
 \partial_t\rho_{12}= (2i\delta-\Gamma_2)\rho_{12} - iA\rho_{02} - iB\rho_{10}
 \label{eq:rho12} 
\end{equation}
\begin{equation}
 \partial_t\rho_{02}= (i(\Delta+\delta)-\gamma_2)\rho_{02} - iB(\rho_{22} - \rho_{00})- iA\rho_{12}
 \label{eq:rho20} 
\end{equation}
\begin{equation}
 \partial_t\rho_{01}= (i(\Delta-\delta)-\gamma_2)\rho_{01} - iA(\rho_{11} - \rho_{00})- iB\rho_{21}
 \label{eq:rho10} 
\end{equation}
\begin{equation}
 \partial_t\rho_{00}= -\gamma\rho_{00} + iB(\rho_{02} - \rho_{20}) + iA(\rho_{01} - \rho_{10})
 \label{eq:rho00} 
\end{equation}
\begin{equation}
 \partial_t\rho_{11}= +\gamma b\rho_{00} +\Gamma(\rho_{22}+\rho_{33}-2\rho_{11}) - iA(\rho_{01} - \rho_{10})
 \label{eq:rho11} 
\end{equation}
\begin{equation}
 \partial_t\rho_{33}= \gamma (1-2b)\rho_{00} +\Gamma(\rho_{11}+\rho_{22}-2\rho_{33})
 \label{eq:rho33} 
\end{equation}
where $b$ is the branching ratio of the decays of $\ket{0}$ into the $\ket{1}, \ket{2}$ subspace and $\gamma, \gamma_2$ are the $T_1$ and $T_2$ rates for the excited state and $A$ and $B$ are proportional to  the laser field amplitude of right ($\sigma^+$) and left ($\sigma^-$) light components times the dipole matrix element. The $\Delta$ we call the one-photon detuning and $\delta$ the two-photon detuning. $\Gamma$ and $\Gamma_2$ are the  $T_1$ and $T_2$ rates for the ground states. 

It is useful to note the discrete symmetries of the EIT equations in the RWA set Eqs.~(\ref{eq:rho12})-(\ref{eq:rho33}). The effective charge conjugation-parity symmetry we call CP \cite{torosov1}. It consists of the transformations $\Delta, \delta \rightarrow -\Delta, -\delta$, along with $\rho \rightarrow \rho^*$  (note: not $\dagger$) and $A,B \rightarrow -A,-B$. This symmetry is distinct from the ${\bf Z}_2$ permutation symmetry $'1' \leftrightarrow '2'$ along with $A \leftrightarrow B$ and $\delta \rightarrow -\delta$ for any fixed $\Delta$. 

The experiment described below uses linearly polarized light so $A=B$. In that case the symmetry CP $\times$ ${\bf Z}_2$ is broken at nonzero $\Delta$, this product symmetry equating the right- (resp. left-) circularly polarized channel of an up-chirp in $\delta$ with the left- (resp. right-) channel of the down chirp in $\delta$. This indicates that the sum of those channels is CP $\times$ ${\bf Z}_2$ symmetric so, at least to leading order for any $\Delta$, we expect no $\delta$-chirp asymmetry in the sum signal. Clearly the individual right- and left- circularly polarized channels are not CP $\times$ ${\bf Z}_2$ symmetric, and below we experimentally measure the $\delta$-chirp asymmetry in these observables. 

The large separation of timescales between the ground state ($\rho_{12}$) and excited state ($\rho_{01}$, $\rho_{02}$) coherences invites one to simplify the above equation via what is colloquially called 'adiabatic elimination'. This will clarify the relevance, if any, for any excited state effects, and make the above symmetries even more obvious, although below all the theory graphics  are prepared using the full set Eqs.~(\ref{eq:rho12})-(\ref{eq:rho33}). For simplicity only, we assume also that the change in the population in $|3>$ during the $\delta$- chirp is negligible. Then, in adiabatic elimination, we have from Eq.~(\ref{eq:rho10}) and Eq.~(\ref{eq:rho20}) that 
\begin{equation}
    \rho_{01} \thickapprox  (\Delta -\delta-i\gamma_2)(A\rho_{11} + B\rho_{21})/D 
\label{eq:adElim01}     
\end{equation}
\begin{equation}
    \rho_{02} \thickapprox (\Delta +\delta-i\gamma_2)(B\rho_{22} + A\rho_{12})/D 
\label{eq:adElim02}     
\end{equation}
where $D = \Delta^2 + \gamma_2^2$. Defining $d = \rho_{11}-\rho_{22}$, we use $\rho_{11} + \rho_{22} \sim const.$ to reduce the homogeneous set 
Eqs.~(\ref{eq:rho12})-(\ref{eq:rho33}) (at $A=B$) to the two pertinent inhomogeneous equations, 
\begin{equation}
\begin{aligned}
  \partial_t\rho_{12} &= \left[i2(1-\frac{A^2}{D})\delta - (\Gamma_2 +\frac{2\gamma_2A^2}{D})\right] \rho_{12}  \\
     & \qquad   -\frac{A^2}{D} (i\delta + \gamma_2)  + i \frac{A^2}{D}  d \Delta
\end{aligned}
\label{eq:adElim12} 
\end{equation}
\begin{equation}
   \partial_t d = -\bigl(\Gamma+\frac{2\gamma_2A^2}{D}\bigr) d 
   +i\frac{2A^2}{D} (\rho_{12}-\rho_{21}) \Delta
\label{eq:adElimd} 
\end{equation}
which clearly have the same symmetries as the original RWA equations, $d$ being CP-even and  ${\bf Z}_2$-odd. Note also since $Im(\rho_{12})$ is CP-odd whereas $Re(\rho_{12})$ is CP even, it is clearly the last term in each of Eq.~(\ref{eq:adElim12}) and Eq.~(\ref{eq:adElimd}) at fixed $\Delta$ that explicitly breaks the effective CP. 

Reducing this 3 real dimension form further, the system can be related to the simplest equation set displaying chirp asymmetry, namely, the two real dimensional basic equations for Leptogenesis.\cite{buchmuller1,buchmuller2}.  In brief that system,  
\begin{equation}
   \frac{{\rm d}N_{B-L}}{{\rm d}Z} = \epsilon {\cal D}(Z)\left(N_N-N_N^{eq}(Z)\right) -W(Z)N_{B-L} \qquad .
\label{eq:leptogenesis1}
\end{equation}
\begin{equation}
   \frac{{\rm d}N_N}{{\rm d}Z} = -\left({\cal D}(Z)+S(Z)\right)\left(N_N-N_N^{eq}(Z)\right)
   \label{eq:leptogenesis}
\end{equation}
is written in terms of a chirp-even quantity $N_N$ indicating the density of particles with no net lepton number that decay into leptons and anti-leptons, and a chirp-odd quantity $N_{B-L}$ representing the net lepton number density. The former basically starts at some equilibrium value $N_N^{eq}$ and the later ( $N_{B-L}$) starts at 0 and grows as $N_N$ deviates substantially from $N_N^{eq}$ during the non-equilibrium expansion/cooling of the Universe. Due to the smallness of $\epsilon$, the $N_{B-L}$  remains small (in comparison with $N_N$) throughout the subsequent evolution. Note that $N_N^{eq}$ also depends on $Z$ ('time').  Here the transient is induced parametrically also through the $Z$-dependence ('time') of the Eqs.~(\ref{eq:leptogenesis1}),(\ref{eq:leptogenesis}) coefficients, $\cal D$, $S$ and $W$. Typically the explicit CP violating parameter $\epsilon$ represents the residual physical effects of a process presumably occurring at a much higher energy (temperature) scale and is thus taken as a constant.

For our system we now make contact with the simplified leptogenesis model above by further reducing the three real dimensional form in Eq.~(\ref{eq:adElim12}) and Eq.~(\ref{eq:adElimd}) to a version of Eqs.~(\ref{eq:leptogenesis1}),(\ref{eq:leptogenesis}). First note that in the $Im(\rho_{12})$ is assumed to be small (its CP odd) and $\delta <<\Delta$ limit, the $Re(\rho_{12})$ doesn't change much as $\delta$ goes through 0. The hierarchy of scales we employ in this simplification is the usual one for atomic EIT systems in that $\gamma_2>>\Gamma, \Gamma_2$ and $A^2/(D\Gamma_2)$ is not too small (though $\frac{A^2}{D\gamma_2}$ is small). Eq.~(\ref{eq:adElim12}) in that limit gives, 
\begin{equation}
   Re(\rho_{12}) \sim \frac{-\gamma_2 A^2}{D\Gamma_2 + 2\gamma_2 A^2}
\label{eq:Rerho12}
\end{equation}
So that, to parallel Eqs.~(\ref{eq:leptogenesis1}),(\ref{eq:leptogenesis}), we eliminate $Re(\rho_{12})$ from Eq.~(\ref{eq:adElimd}) and Eq.~(\ref{eq:adElim12}). In this limit at non-zero $\Delta$ and slow sweep in $\delta$, we have  the equation pair
\begin{equation}
  \partial_t Im(\rho_{12}) \thickapprox \frac{A^2}{D} \Delta (d-d^{eq}) - (\Gamma_2 + \frac{2\gamma_2A^2}{D}) Im(\rho_{12}) 
\label{eq:ImRh12}
\end{equation}
\begin{equation}
   \partial_t d \thickapprox  -(\Gamma+\frac{2A^2}{D}) d + \ldots
\label{eq:dPrime}
\end{equation}
where $d^{eq} \thickapprox -\delta\bigr(\frac{2\gamma_2}{\Delta \Gamma_2}\bigl)$, plays the role of the $N_N^{eq}$ and in Eq.~(\ref{eq:ImRh12}) the $\Delta$ being proportional to the explicit, constant CP-breaking parameter $\epsilon$. Its time dependence through $\delta$ (in $d^{eq}$) during the two-photon chirp drives the $d$ away from 0, ultimately leading to the chirp asymmetry in the difference in the optical transmission of $A$ and $B$. 

The resulting equations Eqs.~(\ref{eq:ImRh12}),(\ref{eq:dPrime}) are structurally similar to Eqs.~(\ref{eq:leptogenesis1}),(\ref{eq:leptogenesis}), enjoying the same transient systematics. Although we have in the forgoing paragraphs focussed on the $\rho_{12}$
and $d$, the actual experimental observables are typically $Im(\rho_{01})$ and $Im(\rho_{02})$ which in the optically thin cell limit are proportional to the absorption of the right- and left- circular components of the light field. In fact, by the forgoing symmetry considerations, the chirp asymmetry is roughly proportional to the difference of these two observables, which, by Eqs.~(\ref{eq:adElim01}), (\ref{eq:adElim02}), is approximately proportional to the CP $\times $ ${\bf Z}_2$ odd quantities $\Delta d$ and $Im(\rho_{12})$. 

Since our goal here is to connect the EIT Chirp phenomenology with the behavior common to systems with broken discrete symmetries that are driven off equilibrium we have described the connection with the simplest model of leptogenesis for pedagogic completeness only. Instead for the comparison with experiment we solve the full set Eqs.~(\ref{eq:rho12})-(\ref{eq:rho33}) and use that numerical solution in all the theory graphs below. 

To model the $^{87}$Rb-buffer gas cells optical response, we fix $A=B$ and take $\delta$ to be much smaller than $\Delta$. In our buffer gas cell the  $\Gamma$ and $\Gamma_2$ are also quite small, and physically are fixed by the diffusion of atoms in and out of the beam and residual magnetic field variances. In the simulation we take $\gamma=1$ to set the overall scale, and we solve these differential equations for a particular chirp speed in $\delta$. The resulting $\rho$ are then fastened into observables such as the absorption coefficients of the right- and left- circular polarization components of the light field ($A$Im$(\rho_{10})$ and $B$Im$(\rho_{20})$ respectively) and their sum.  

On a related qualitative note, theory and experiment both indicate the well-known fact that the lineshape asymmetry in the individual polarization components is much larger than in the sum. Also, both experiment and theory at high chirp speed (sometimes referred to as the non-adiabatic regime\cite{jin, chen_yf, grewal} ) show an oscillatory component of the response in each component and, to a lesser degree, in their sum. We limit our investigations here to chirp rates below the onset of significant oscillation. The individual circular polarization channels do indeed have a much more asymmetric lineshape although their sum is rather more symmetrical (see Fig.\ref{fig:signals}). In the experiments at higher chirp speeds (starting at about 50 Hz) substantial oscillatory ringing is seen in both the individual circular polarization channels and the sum. The onset of this ringing appears to be at about the set of parameters for most of the experimental protocols we employ, and thus we expect any differential proximity to the non-adiabatic regime to not be relevant to the observed chirp asymmetry systematics we study below.  

The breaking of the effective CP symmetry that arises in the RWA from a fixed, non-zero one-photon detuning leads to marked differences in magnitude and lineshape between up- and down- two-photon detuning chirps.  For definiteness we compare the peak optical response during the up- and down- chirps. Throughout the "asymmetry" we tabulate and graph is the ratio of the peak optical response during an up-chirp minus that during a down-chirp all divided by their sum \cite{commons}. 

\section{Experiment}
\begin{figure}
  \centering
  \includegraphics[width=0.50\linewidth]{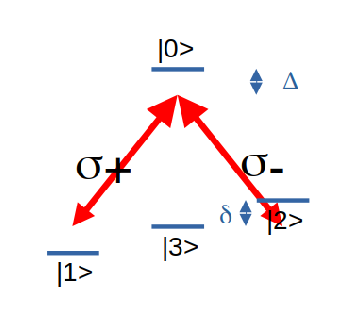} 
  \includegraphics[width=1.00\linewidth]{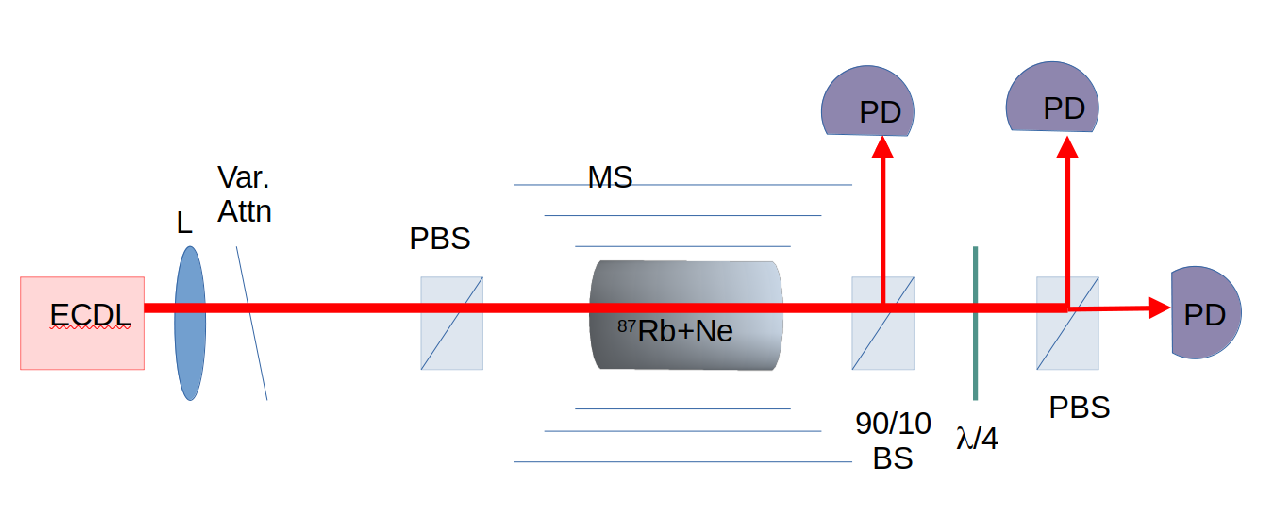}
  \caption{(a) The basic level diagram for the theory model and,
  (b) The minimal Hanle/Zeeman EIT arrangement: ECLD = extended cavity diode laser, L = lens (to expand the beam), VA = variable attenuator, BS = nonpolarizing beam splitter, PBS = polarized beam splitter, $^{87}$Rb+Ne = vapor cell, MS = triply magnetically shielded cavity, PD = photodetectors, $\lambda/4$ = zero-order quarter wave plate. The solenoid and gradient loop around the cell are not shown.}
  \label{fig:setup}
\end{figure}

We employ the usual Hanle resonance configuration shown in Fig.\ref{fig:setup}b. Laser light from an extended cavity diode laser (NewFocus Vortex ECDL) tuned to the D1 (795nm) resonance in Rubidium is expanded and (linearly) polarized before traversing a 5cm long pyrex vapor cell. We also performed the same measurements with a temperature-tuned free-running 795nm laser diode and note here that the resulting resonance systematics and chirp asymmetry were easy to discern and quite similar to that of the ECDL, in spite of its expected $\sim$50MHz laser phase noise compared to the ECDL ($<1$ MHz). All the data described below was collected exclusively with the NewFocus Vortex ECDL. Another simplification afforded by our use of the Hanle-Zeeman resonance was that only commodity small bandwidth ($<$5MHz) optical detectors (compared to the high speed, large gain-bandwidth product detectors in the experiment of Ref. \cite{commons}) were necessary for all light power measurements. 

Throughout this experiment the vapor cell contained a mixture of 2 Torr of Neon and isotopically enriched $^{87}Rb$ (Opthos). It is housed at the center of a triply layered mu-metal magnetically shielded cylinder with end caps that have a 2.5cm hole for optical access. Inside the innermost mu-metal shield is a solenoid (~23G/A) and a pair of 'gradient' loops on each end of the innermost cavity. The 'gradient' loops are connected in an anti-Helmholtz sense. The entire shielded assembly+cell+solenoid is in a thermostatically controlled box, heated with a low magnetic signature heater (HTD Heat Trace, Inc.) to 56$^o$C.  
Light emerging from the other side of the cell passes first through a nonpolarizing 90-10 beam cube, with the 10\% being recorded as our 'sum' signal and the remaining 90\% then sent through a zero order quarter wave plate followed by a polarizing beam cube oriented so that from the output ports of this final beam cube emerge the left- ($\sigma^-$) and right- ($\sigma^+$) polarized components of the lightfield. The three light fields ('sum', "$\sigma^+$" and "$\sigma^-$") power is measured on the $\sim$ .5x.5cm amplified photodiodes whose signals are simultaneously digitized and recorded.

While the ECDL is tuned and weakly locked (fixed $\Delta$) within the doppler+collisionally broadened $F=2 \rightarrow F'=1$  $^{87}$Rb transition, a symmetric triangle waveform (provided by an SRS DS 345) with zero offset and an amplitude of 0.35V is impressed across a series combination of a 73K$\Omega$ resistor and the solenoid inside the shields, creating our chirp in the two-photon detuning $\delta$. As the total magnetic field in the vapor along the beam sweeps through zero the level-crossing degeneracy leads to an increase in light transmission called the Hanle/Zeeman EIT resonance. 

\begin{figure}
  \centering
  \includegraphics[width=1.0 \linewidth]{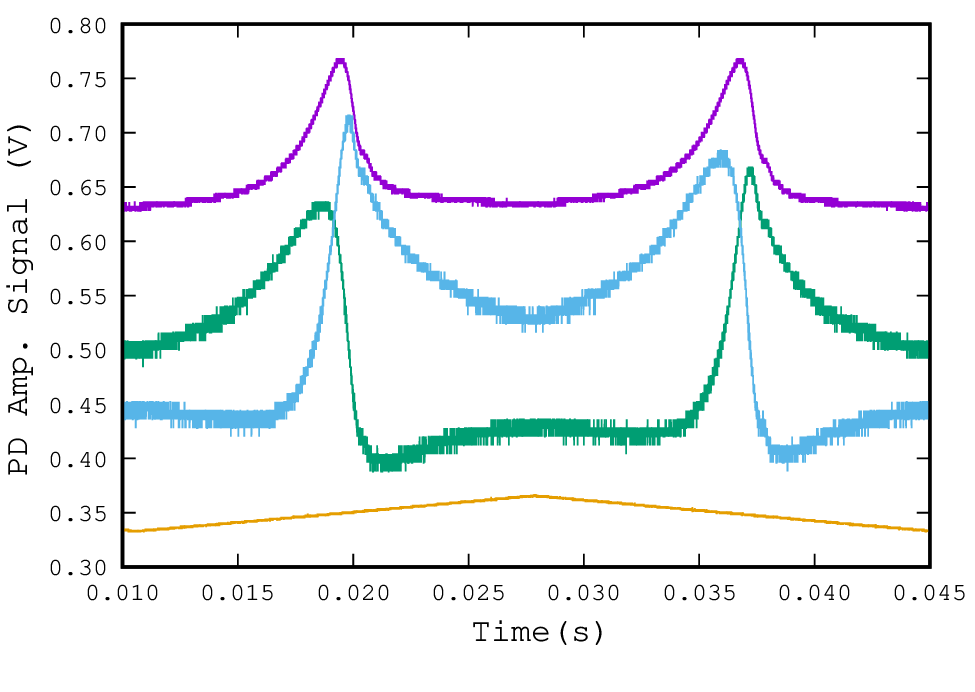} 
  \caption{Typical experimental signals, (top trace) the sum shown along with the $\sigma^+$  and $\sigma^-$ channels as a function of the current in the solenoid (bottom trace) all at a chirp speed of about 0.16 MHz/s and optical power 230$\mu$W and a spot size of about 1.5cm$^2$. Chirp asymmetry is plainly visible (and of opposite sign between polarizations) in both channels though the sum is chirp symmetric (i.e. comparing response in up chirp at $\delta$ to the down chirp at $-\delta$). Signals offset vertically for ease of viewing.}
  \label{fig:signals}
\end{figure}

A typical trace of all three channels displaying these features is shown in Fig.\ref{fig:signals}. There the traces were taken at a very slow sweep (about 2Hz). Differences in the peak intensities of the peak in the right-circular up-chirp and left-circular down-chirp are primarily due to differences in photodiode amplifier gain. Since the asymmetry is computed by comparing the optical response in up- and down-chirp in the same channel, any differences in photodiode amplifier gain do not materially enter into the asymmetry computed and graphed below. Finally we note that the Hanle EIT width depends on optical power (more below), and in this experiment the extrapolated zero optical power limit of the EIT (full-)linewidth for this cell and shield configuration was $\sim$400 Hz. 

\section{Discussion} 
As a demonstration and application of our emerging understanding of chirp asymmetry systematics, our experimental protocol makes use of the parametric malleability of the Hanle EIT resonances. 
In quantitative comparisons with theory throughout we have numerically evaluated the chirped response using the full 4-level theory Eqs.~(\ref{eq:rho12})-(\ref{eq:rho33}). That evaluation includes integrating while varying $\Delta$ chosen randomly from a Boltzmann distribution, the one-photon detunings experienced by an atom diffusing through the beam. An additional random contribution to the one-photon detuning is also included in the integration in an attempt to reproduce any effects from laser phase noise. 

\begin{figure}
  \centering
  \includegraphics[width=1.0\linewidth]{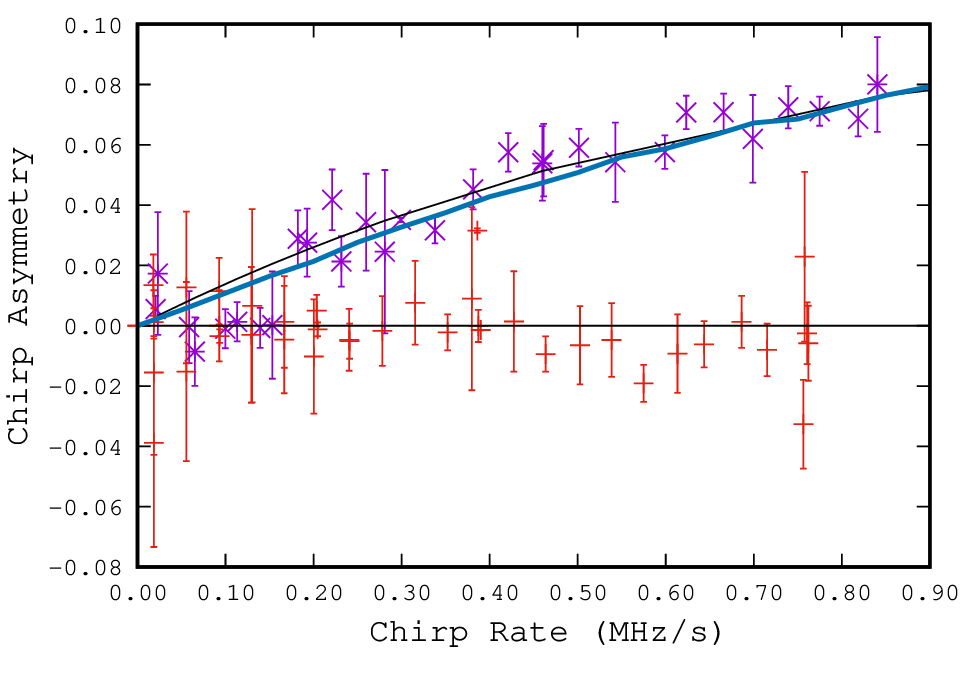} 
  \caption{A plot of the experimentally measured chirp asymmetry in $^{87}$Rb F=2$\rightarrow$ F'=1 transition.  Purple data points are from the transmitted $\sigma^+$ and red points are from the EIT resonance both at the same total beam power ($\sim$ 230 $\mu$W). The $\sigma^+$ data shown was taken at a one-photon detuning of roughly -200MHz, but the red data is at this and different one-photon detunings. The thick dark blue line is the theoretical chirp asymmetry from integrating Eqs.~(\ref{eq:rho12})-(\ref{eq:rho33}), whereas the light grey line is simply a two parameter ($a$, $b$) fit of the experimental data in purple to the functional form $a(1-e^{-bx})$ for comparison. Error bars (vertical) are ensemble deviations from multiple measurements.} 
  \label{fig:expSum}
\end{figure}

Again, the ratio of the peak optical response during an up-chirp minus that during a down-chirp all divided by their sum is what we label "Chirp Asymmetry" although it is but a single $CP$-odd observable. On general theoretical grounds we expect any $CP$-odd observables in a system in which the symmetry is explicitly broken to have the characteristic dependence on chirp speed as detailed in Ref.\cite{commons}, namely, an initial linear growth of the asymmetry with chirp speed followed by saturation (at a value not necessarily $\pm$1). Likewise we expect the $CP$-even observable (the sum channel) to show no appreciable chirp asymmetry at any chirp speed or any value of the $CP$-breaking parameter (here one photon detuning, $\Delta$). 

The experimental summary of data in Fig.\ref{fig:expSum} largely bears out theory expectations. The 'sum' signal (red data points) is $CP$ even and shown are data taken at various one-photon detunings. 
A linear fit to the 'sum' data give both a slope and intercept statistically identical to 0.  The $\sigma^{+}$ 
data (purple data points) on the other hand was taken at a fixed one-photon detuning of -200 MHz. Shown alongside that data is an integration of the Eqs.~(\ref{eq:rho12})-(\ref{eq:rho33}) (dark blue line) as described earlier and using typical experimental values.  

In that figure is also a thin grey line. It is a fit of these data to an naive two-parameter relaxation curve and not based on theory. It indicates a  asymptotic (large chirp rate) asymmetry of just under 12 percent. In actual experiment and the full physical theory of this system the asymptotic value is less meaningful because beyond a certain chirp speeds the response becomes oscillatory (the so-called "non-adiabatic" regime). An oscillatory regime is not present in the dimensionally reduced "leptogenesis"-like models described in the later part of the theory section as that model has only populations. The data and theory shown in Fig.\ref{fig:expSum} ends at the onset of that non-adiabatic regime. 

\begin{figure}
  \centering
  \includegraphics[width=0.8\linewidth]{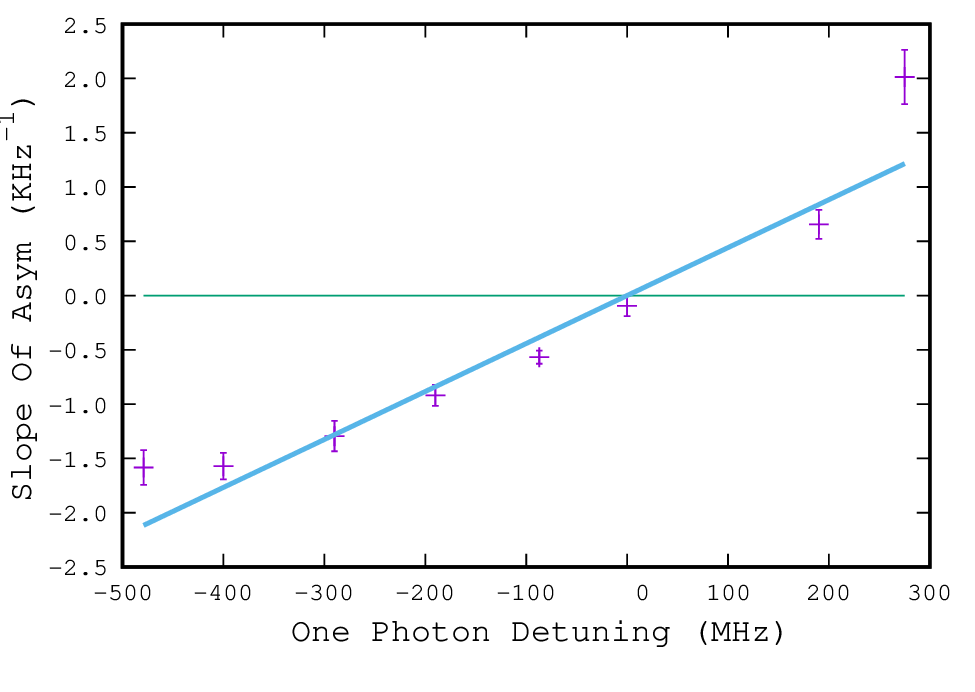} 
  \caption{ A plot of the experimentally measured chirp asymmetry slopes at small sweep speeds as one changes the one-photon detuning. Line shown is a fit of the data between -350 to 250 MHz via a single parameter (a slope, no intercept). This functional form is expected theoretically at small one-photon detunings, the data indicating significant deviations beyond that region. For the data points shown the error in detuning is estimated but the error in chirp asymmetry is an ensemble deviation from multiple measurements. } 
  \label{fig:expSlopes}
\end{figure}

Next we test the theory contention that the one-photon detuning is a $CP$-breaking parameter. By using different one-photon detunings while solving equations Eqs.~(\ref{eq:rho12})-(\ref{eq:rho33}), evaluating the theory indicates a roughly linear dependence of the $\delta$-chirp asymmetry slope near $\Delta = 0$.  In Fig. \ref{fig:expSlopes} we have plotted the experimentally measured asymmetry slope near zero chirp speed for $\sigma^+$ as a function of the laser's one-photon detuning along with a one-parameter (intercept zero) fit line. The data suggests a nearly linear dependence on the one-photon detuning over most of the experimentally accessible range, with a zero asymmetry point statistically consistent with zero one-photon detuning. 

Finally, asymmetry being dimensionless implies dependence on the chirp rate is scaled to the resonance's linewidth. That is, increasing the linewidth is expected to have the same effect as proportionally reducing the chirp rate. We use small currents in the gradient coil pair to controllably inhomogeneously broaden the Hanle/EIT resonance. We also experimentally study the effect on chirp asymmetry of (optical-)power broadening the resonance, thereby determining the role of homogeneous line broadening plays in modulating the chirp asymmetry. For each of these experimental tests tabulated below we make quantitative and qualitative contact with theory, providing a rather rigorous test of our model and understanding of chirp asymmetry resulting from a broken discrete symmetry. 

\begin{figure}
  \centering
  \includegraphics[width=0.8\linewidth]{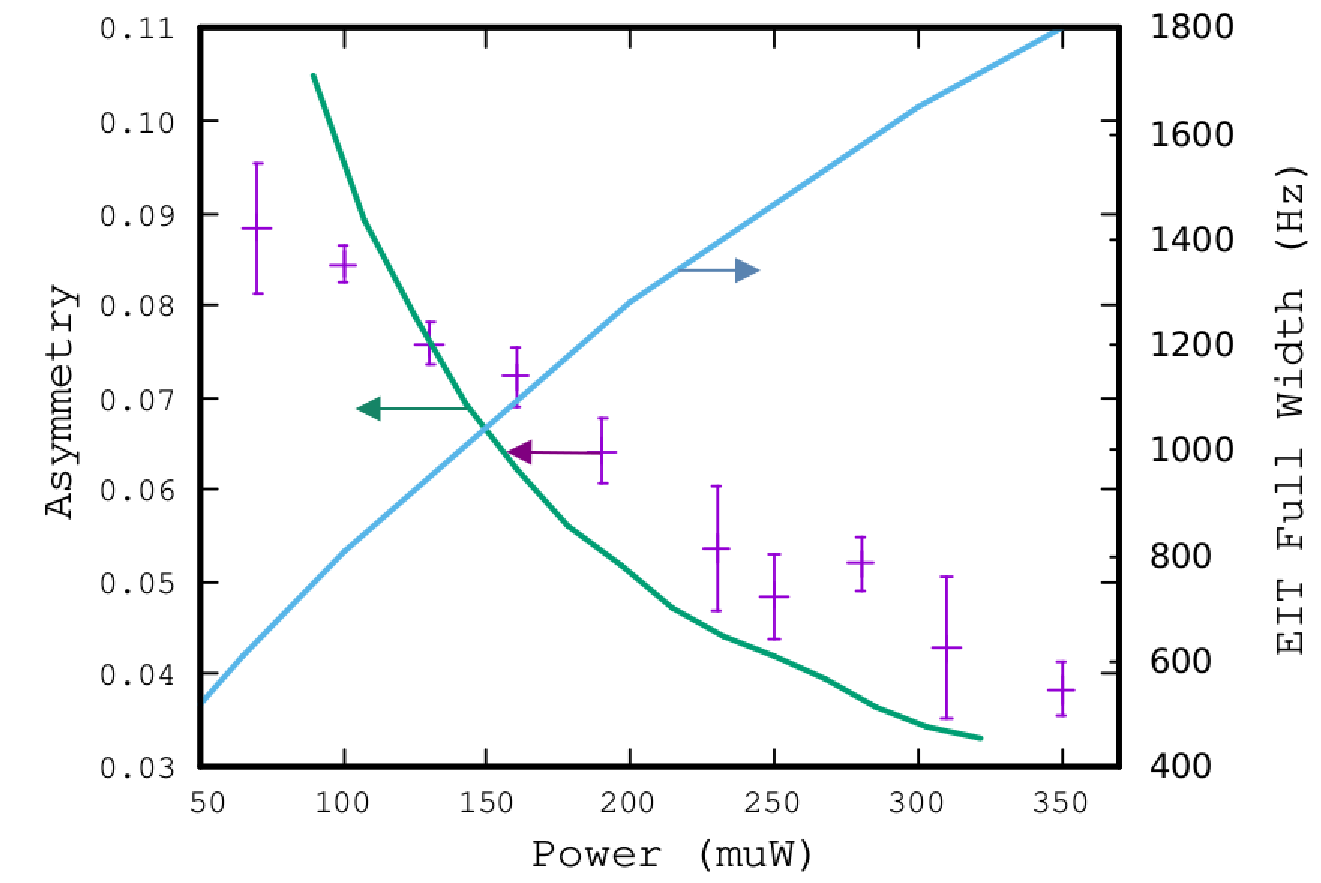} 
  \caption{The optical power dependence of the experimental chirp asymmetry (axis left) in the $\sigma^+$ signal, along with power-broadened EIT ('sum') Width from experiment (axis right). The green line is from integrating the transient chirp using Eqs.~(\ref{eq:rho12})-(\ref{eq:rho33}) for a single power (flat-top beam).  For the experimental points shown the error in the measured beam power is estimated but the error in chirp asymmetry is an ensemble deviation from multiple measurements. The most likely explanation consistent with the apparent deviation between the theory curve (green) and the data points (purple) is that our beam differs significantly from a flat-top beam.}
  \label{fig:expWidth}
\end{figure}

Fig. \ref{fig:expWidth} is a graph of the measured Hanle/EIT linewidth as a function of the measured optical power of the linearly polarized lightfield. The trend and magnitude of the effect there clearly indicates that as the EIT linewidth (blue line, scale right) is broadened by a factor of three the chirp asymmetry (purple data points, scale left) is reduced by nearly the same factor. The theory curve (green, scale left) was computed by integrating Eqs.~(\ref{eq:rho12})-(\ref{eq:rho33}) at a detuning of -150 MHz and a sweep rate of $\sim$ 45 Hz (near the end of Fig. 3, ${\hat x}$-axis) but graphed with power via a  single positive offset and power scale factor. The later was necessary to account for the vagaries of the beam shape and power to the value of the $'A'$ (=$'B'$) field amplitude in the numerical evaluation of the theory. The former (offset in power scale) being positive and small (relative to the axis scale) indicates that, plausibly, there are other contributions to the broadening of the transition that persist at zero power. Note that the theory curve being steeper than the experimental data is likely due to the fact that our beam cross-section is not intensity-flat (the simulation being computed at one intensity, essentially, treating the beam as a flat-top beam). 

\begin{figure}
  \centering
5  \includegraphics[width=0.8\linewidth]{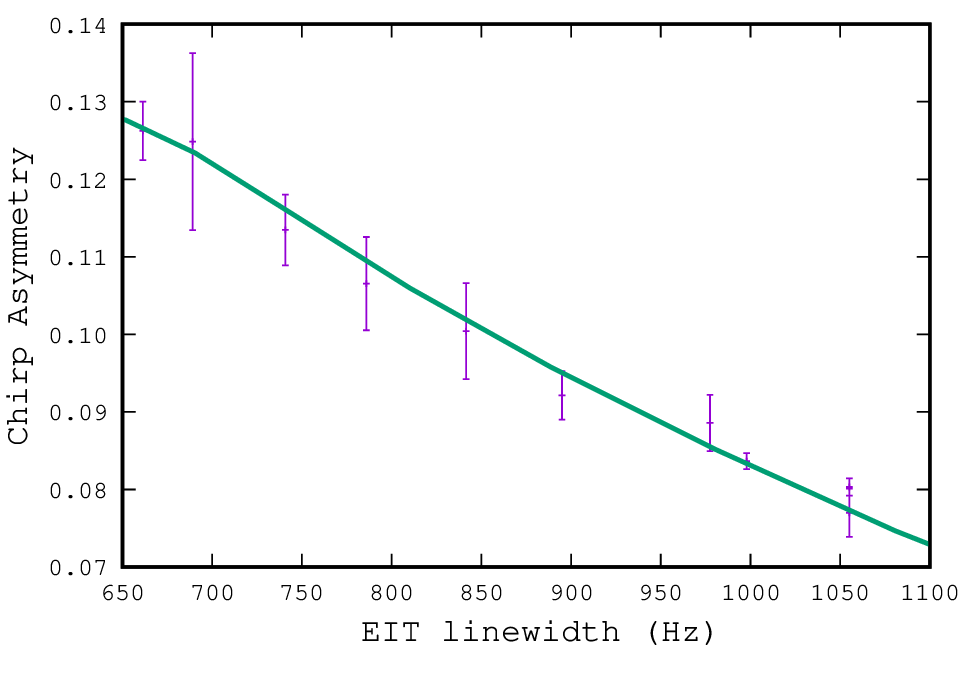} 
  \caption{A plot of the experimentally measured chirp asymmetry in the $\sigma^+$ signal as a function of gradient-induced EIT ('sum') width. The measured EIT width change is very nearly linear in the current in the gradient loop pair (not shown). Error bars (vertical) again represent ensemble deviations from multiple measurements. The line (green) is a theory curve computed using a shifted base theory curve (fuller explanation in text)} 
  \label{fig:expWidthGradient}
\end{figure}

We can also cause controlled growth of the width of the EIT resonance by inhomogeneous broadening using small currents in the gradient coils. The resulting data and theory curve are assembled in Fig. \ref{fig:expWidthGradient}. These data were taken at a detuning closer to -200MHz, and a power of under 100$\mu$W and a fixed chirp rate of 45 Hz (close to 0.8 MHz/$\mu$S in $\delta$'s rate). For the associated theory curve we first generated a single 'reference' theory trace of \{$\sigma^+, \sigma^-$\} in $\delta$ by solving Eqs.~(\ref{eq:rho12})-(\ref{eq:rho33}) under the aforementioned parameters yielding a chirp asymmetry close to 0.13. Assuming the vapor was optically thin, to model the effect of the gradient coils we then simply shifted and co-added many copies of this 'reference' trace in $\delta$ as the shift was finely scanned across a window (centered about $\delta=0$) $[-W,W]$ for a given $W$. Then the chirp asymmetry (in $\sigma^+$) and EIT width (from $\sigma^+ + \sigma^-$) were computed for the co-added result for each $W$. Plotting then the asymmetry against the EIT width of this synthetic signal leads to the green curve in Fig. \ref{fig:expWidthGradient}. 

Oftentimes inhomogeneous broadening is incorporated into numerical evaluation of a model by simply plotting the signal versus $\gamma_2$. Our rationale for doing the above windowed co-adding was that it was, we felt, more representative of the actual physical cause of inhomogeneous broadening in our optically thin vapor cell at low irradiance because exciting the gradient coils led to a nearly linear gradient of two-photon detuning along the optical axis of the cell.

This study suggests metrological consequences in bandwidth and modulation scheme dependence due to broken discrete symmetries in quantum optical systems, of particular importance in multi-photon processes. Though it is hard to posit an example with a numerical value, for EIT employed in time/frequency metrology, this work indicates polarization 'impurity' in what was regarded as a $CP$-even (like 'sum') signal will lead to modulation rate/depth dependence that might be otherwise hard to simply ascribe to a 'lineshape' or 'modulation method/parameter' issue. 

Work is underway to find different theoretical descriptions of this general phenomenon as a way to connect it to other non-equilibrium phenomenon. Effects of cross modulation from other (non $CP$) broken symmetries, as well as the effect caused by higher order non-linear optical processes in such systems may be fruitful avenues for further elaboration. 

To conclude, we have tested the chirp asymmetry theory model experimentally by using the Hanle Zeeman EIT resonance in a $^{87}Rb$ vapor in a buffer gas cell. By changing the one-photon detuning and/or subjecting the cell to a longitudinal field gradient and using optical power broadening we were able to verify that the one-photon detuning is a $CP$-odd perturbation and affects that increased the  two-photon width reduced the measured asymmetry at fixed chirp rate. Broadening contributions both homogeneous and  inhomogeneous lead to a reduction in the chirp asymmetry. Data suggests that the product of the asymmetry and the total resonance width varying only slowly with chirp parameters,  indicating the likelihood (at least in simple two-photon transitions) of a single relevant timescale parameterizing how far from equilibrium the system is driven during a chirp (see for example the 'chirp parameter' defined in Ref. \cite{ernst}).

\section*{Funding}
MC acknowledges National Science Foundation support under NSF-DMR-2226956, as well as partial support from ITAMP early in the drafting of the manuscript and support from the Quantum Technology Center (QTC) at the University of Maryland where this manuscript was completed. JG, KG, NO, VT and DT acknowledge support provided by the YSU Ann Seimon endowment for undergraduate research in physics. 

\section*{Disclosure} 
The authors declare no conflicts of interest.

\bibliography{refs}

\end{document}